\documentclass{elsart}
\usepackage{epsfig}
\usepackage{amssymb}
\begin{document}

\begin{frontmatter}

\title{Quantum Mechanics without Nonlocality}

\author{N L Chuprikov}

\address{Tomsk State Pedagogical University, 634041, Tomsk, Russia}

\begin{abstract}
We argue that quantum nonlocality of entangled states is not an actual phenomenon.
It appears in quantum mechanics as a consequence of the inconsistency of its
superposition principle with the corpuscular properties of a quantum particle. In
the existing form, this principle does not distinguish between macroscopically
distinct states of a particle and their superpositions: it implies introducing
observables for a particle, even if it is in an entangled state. However, a particle
cannot take part simultaneously in two or more alternative macroscopically distinct
sub-processes. Thus, calculating the expectation values of the one-particle's
observables, for entangled states, is physically meaningless: Born's formula is not
applicable to such states. The same concerns the entangled states of compound
quantum systems. In the {\it existing} quantum mechanics, introducing Bell's
inequalities is fully legal. However, these inequalities imply averaging over an
entangled state, and, hence, they have no basis for their clear physical
interpretation. Experiments to confirm the violation of Bell's inequalities do not
prove the existence of nonlocality in microcosm. They confirm only that correlations
introduced in the existing theory of entangled states have no physical sense, for
they contradict special relativity.

\end{abstract}

\end{frontmatter}

\newcommand{\ppp}{\mbox{\hspace{5mm}}}
\newcommand{\ooo}{\mbox{\hspace{3mm}}}
\newcommand{\ooa}{\mbox{\hspace{1mm}}}

\section{Introduction}

As is well known, quantum nonlocality of entangled states is, perhaps, the most
intriguing question to arise in the modern physics. Quantum theory \cite{Ens,Bel}
and experiment \cite{Asp} say that in the case of entangled states there should be
nonzero correlations between two space-like separated events. From the practical
point of view, this property of entangled states is one of the most desirable
findings of quantum mechanics. For it opens unusual perspectives in developing the
various forms of information technology (see, e.g., \cite{Ber}).

However, despite these opportunities, the phenomenon of quantum nonlocality has
remained as an undesirable "guest" in the modern physics. As is known (see, e.g.,
\cite{No1,No2,No3,Wis}), Einstein waged a relentless struggle against this
prediction of quantum mechanics. For it contradicts the principles of special
relativity, and, as a result, it violates the continuity in science.

To reconcile quantum mechanics with special relativity, Heisenberg was perhaps first
(see \cite{Hei} and \cite{No1}) who stated that quantum nonlocality "... is not in
conflict with the postulates of the theory of relativity." This idea was later
developed by Bell \cite{Bee} (see also a quote in \cite{No1}): having proved his
famous theorem, Bell was disturbed with his own result. In order to smooth the
contradiction between quantum mechanics and special relativity, Bell suggested that
nonzero correlations between space-like separated events do not at all mean the
existence of a superluminal signalling between such two events (see also
\cite{Hei}). As is now accepted in all the so-called non-signalling theories (see,
e.g., \cite{Gis}), quantum nonlocality is an {\it uncontrollable phenomenon}
\cite{Shi}.

However, one should recognize that such terms as 'nonzero non-signalling
correlations' and 'uncontrollable phenomenon' are too ambiguous. They do not explain
the phenomenon of quantum nonlocality (see also \cite{Hor}). These terms themselves
need explanation.

It should be stressed that Bell himself considered the 'non-signalling' explanation
as a hard choice. He wrote (quoted from \cite{No1}): "Do we then have to fall back
on 'no signaling faster than light' as the expression of the fundamental causal
structure of contemporary theoretical physics? That is hard for me to accept. For
one thing we have lost the idea that correlations can be explained, or at least this
idea awaits reformulation. More importantly, the 'no signaling' notion rests on
concepts which are desperately vague, or vaguely applicable."  So, Bell says in fact
that the 'non-signalling' explanation of nonzero correlations between space-like
separated events makes meaningless the very notion of 'correlations'.

We agree entirely with these doubts. All the known 'no-signalling' explanations are
based on the implicit assumption that the principles of quantum mechanics are
mutually consistent. However, this is not the case. To prove this statement, we
address the quantum problem of a completed scattering of a particle on a static
one-dimensional (1D) potential barrier (see \cite{Ch1}). In this case the analysis
of quantum nonlocality is essentially simpler than for compound systems considered
in the 'no-go' theorems, where there are fundamental problems associated with
multipartite quantum measurements (see, e.g. \cite{Smo}).

\section{On the phenomenon of quantum nonlocality in the existing model of a 1D
completed scattering}

As is known (see reviews \cite{Ha2,La1,Olk,Ste,Mu0,Nu0,Ol3}, during the last three
decades this quantum phenomenon have been in the focus of the intensive debate on
the so-called tunneling time problem, without reaching any consensus.

It should be noted that solving this problem have not been aimed to prove the
existence of quantum nonlocality. At the same time the latter has arisen in all the
existing approaches to the tunneling time problem. The well-known group- and
dwell-time concepts \cite{Ha2,Ha1,Ja1,Le1,But} are not exceptions. Like other
concepts they lead to the unrealistic tunneling times for a scattering particle. As
it has turned out, for a transmitted particle the group and dwell times may be
anomalously short or even negative by value.

It is important to stress that all these approaches, like the 'no-go' theorems, do
not doubt a consistency of the quantum-mechanical principles. And, like the 'no-go'
theorems, they in fact prove that nonlocality is an inherent property of
conventional quantum mechanics.

However, just the main peculiarity of the existing quantum-mechanical model of a 1D
completed scattering is that it is inconsistent (in details, this question is
studied in \cite{Ch1}). This fact can be demonstrated, for example, in the case of
the Bohmian mechanics. Indeed, the Bohmian model of the process (see \cite{Le2})
predicts that the fate of an incident particle (to be transmitted or to be reflected
by the barrier) depends on the coordinate of its starting point. However, this
property is evident to contradict the main principles of quantum mechanics, since a
starting particle should have both the possibilities, irrespective of the location
of its starting point.

It is evident that this peculiarity of the existing Bohmian model is closely
connected to spatial nonlocality. Indeed, the position of a critical spatial point
to separate the starting regions of to-be-transmitted and to-be-reflected particles
depends on the shape of the potential barrier (though the barrier is located at a
considerable distance from the particle's source). Thus, the "causal" trajectories
of transmitted and reflected particles, introduced in the Bohmian mechanics, are, in
fact, non-causal: they are ill-defined.

So, in the Bohmian model of a 1D completed scattering, nonlocality and inconsistency
accompany each other. However, this situation is common for all the known approaches
to deal with a 1D completed scattering (see \cite{Ch1} and references therein). Our
analysis shows that quantum nonlocality to arise in the existing model of a 1D
completed scattering results from the inconsistency of the superposition principle
with the corpuscular properties of a particle.

\section{A 1D completed scattering as an entanglement of transmission and
reflection}

In \cite{Ch1}) we have presented a new model of a 1D completed scattering for a
particle impinging a symmetrical potential barrier from the left. We show that, for
a given potential and initial state of a particle, the (full) wave function
$\psi_{full}(x;E)$ to describe this process, for a particle with a given energy $E$,
can be uniquely presented in the form,
\[\psi_{full}(x;E)=\psi_{ref}(x;E)+\psi_{tr}(x;E),\]
where $\psi_{ref}(x;E)$ and $\psi_{tr}(x;E)$ are solutions to the stationary
Schr\"odinger equation. They are such that allow us to retrace the time evolution of
the (to-be-)transmitted and (to-be-)reflected {\it subensembles} of particles at all
stages of scattering.

As it has been shown in \cite{Ch1}, for any symmetric potential, $\psi_{ref}(x;E)$
is an odd function with respect to the midpoint $x_c$ of the barrier region; i.e.,
for any value of $E$ we have $\psi_{ref}(x_c;E)=0$. This means that particles
impinging the barrier from the left do not enter the region $x>x_c$.

Let the wave function to describe the subensemble of such particles be denoted
$\tilde{\psi}_{ref}(x;E)$. Then
\begin{eqnarray} \label{1}
\tilde{\psi}_{ref}(x;E)\equiv \psi_{ref}(x;E)\ooo for\ooo  x\le x_c; \ppp
\tilde{\psi}_{ref}(x;E)\equiv 0 \ooo for\ooo x\ge x_c.
\end{eqnarray}

Correspondingly, the function $\tilde{\psi}_{tr}(x;E)$ -
\[\tilde{\psi}_{tr}(x;E)= \psi_{full}(x;E)-\tilde{\psi}_{ref}(x;E)\] - which can be
presented also as
\begin{eqnarray} \label{2}
\tilde{\psi}_{tr}(x;E)\equiv \psi_{tr}(x;E)\ooa for\ooa  x\le x_c; \ooo
\tilde{\psi}_{tr}(x;E)\equiv \psi_{full}(x;E) \ooa for\ooa x\ge x_c,
\end{eqnarray}
describes the subensemble of particles with energy $E$, which impinges the barrier
from the left and then passes through the barrier, without reflection and without
violating the continuity equation at the midpoint $x_c$. This property of
$\tilde{\psi}_{tr}(x;E)$ results from the fact that the solutions $\psi_{tr}(x;E)$
and $\psi_{full}(x;E)$ have the same probability current density and, besides,
$\psi_{tr}(x_c;E)=\psi_{full}(x_c;E).$

It is evident that the wave packets $\tilde{\psi}_{tr}(x,t)$ and
$\tilde{\psi}_{ref}(x,t)$ formed from $\tilde{\psi}_{tr}(x;E)$ and
$\tilde{\psi}_{ref}(x;e)$, respectively, obey the time-dependent Schr\"odinger
equation everywhere except for the point $x_c$. However, the continuity equation is
not violated at this point. So that both the wave packets, being everywhere
continuous at any instant of time, evolve in time with constant norms.

By our approach, namely $\tilde{\psi}_{tr}(x,t)$ and $\tilde{\psi}_{ref}(x,t)$ (each
possesses one incoming and one outgoing packet) describe the time evolution of the
(to-be-)transmitted and (to-be-)reflected subensembles of particles at all stages of
scattering. In this case,
\[\psi_{full}(x,t)=\psi_{tr}(x,t)+\psi_{ref}(x,t)=\tilde{\psi}_{tr}(x,t)+
\tilde{\psi}_{ref}(x,t),\] where $\psi_{tr}(x,t)$ and $\psi_{ref}(x,t)$ are the wave
packets formed from the solutions $\psi_{ref}(x;E)$ and $\psi_{tr}(x;E).$

Thus, the superposition of two solutions to the Schr\"odinger equation,
$\psi_{tr}(x,t)$ and $\psi_{ref}(x,t)$, is equivalent to that of
$\tilde{\psi}_{tr}(x,t)$ and $\tilde{\psi}_{ref}(x,t)$ to describe the
sub-processes, transmission and reflection. Since these sub-processes are
macroscopically distinct at the final stage of a 1D completed scattering, the state
of a particle to take a part in a 1D completed scattering should be considered as an
entangled one. This means, in particular, that the quantum nonlocality to arise in
the previous model of a 1D completed scattering is just that inherent to entangled
states.

At this point it is important to stress that transmission and reflection are not
independent quantum processes: they are two alternatives to arise for a particle in
the same scattering problem. So that it is not surprising that
$\tilde{\psi}_{tr}(x,t)$ and $\tilde{\psi}_{ref}(x,t)$ to describe these
sub-processes are not {\it independent} solutions to the Schr\"odinger equation. As
is seen, they can be considered only together, as constituent parts of the same
entangled state $\psi_{full}(x,t)$.

The study of temporal aspects of a 1D completed scattering, carried out on the basis
of $\tilde{\psi}_{tr}(x,t)$ and $\tilde{\psi}_{ref}(x,t)$ (see \cite{Ch1}), has
shown that the behavior of both the sub-processes does not exhibit quantum
nonlocality. They evolve in time without superluminal velocities.

Besides, it has been stated that only the dwell time can be considered as a measure
of the time spent, on the average, by particles of either subensemble in the barrier
region. This characteristic time can be measured with the help of the Larmor clock,
without demolishing the scattering process. It is important to stress here that the
probability fields for transmission and reflection, being superimposed, do not
influence each other.

As regards the group and other characteristic times (which cannot be measured with
the Larmor clock), they seem to have no physical sense when a particle is in
entangled state. By our approach, none point of the wave packet can be used as a
representative of a particle (in an entangled state) in timing its motion in the
barrier region. It says that it is impossible, with the help of the Larmor clock, to
track the motion of any point of a moving wave packet.

One remark should be made also in regard to the Bohmian quantum mechanics. As is
seen, our model in fact implies the introduction of two individual sets of the
causal trajectories for a particle, both for transmission and reflection. In this
case, in a full accordance with the quantum-mechanical principles, each starting
particle has two possibilities, irrespective of the coordinate of its starting
point.

\section{The wave-particle duality and superposition-decomposition principle for
entangled states}

By our approach, the above decomposition of the entangled one-particle's state is
the only way to explain the properties of a 1D completed scattering, since all
one-particle's observables can be introduced only for unentangled one-particle's
states, i.e., for transmission and reflection. As regards any entangled state of a
scattering particle, introduction of observables, which would be common for the
transmitted and reflected subensembles of particles, has no physical sense.

Indeed, a particle, as an indivisible object, cannot simultaneously take part in two
(or several) macroscopically distinct processes. This means that its 'quantum
trajectory', which must be non-divaricate, can be presented only by an unentangled
time-dependent state. Only such state may serve as a counterpart to the classical
one-particle trajectory. All quantum-mechanical rules (including Born's
interpretation of the squared modulus of the wave function as well as Born's rule of
calculating the expectation values of observables) have physical sense only for
unentangled ("non-dendritic") time-dependent states.

This also concerns calculating the correlations between two events for compound
quantum systems. As is known, such calculations imply averaging over the state of a
system. By our approach, such averaging is meaningful only if both these events
belong to the same "non-dendritic quantum many-particle trajectory", i.e., if the
system is in an unentangled time-dependent state.

So, the quantum-mechanical superposition principle must distinguish, on the
conceptual level, macroscopically distinct states and their superpositions. In other
words, it should distinguish unentangled and entangled quantum states. All
observables can be introduced only for unentangled states.

By our approach, if some pure entangled one-particle's state implies the motion of a
particle along a macroscopically dendritic way, hence we deal with an entangled
state. And, in order to explain the motion of a particle along this way, we have to
decompose the entangled state into unentangled (elementary) ones which do not
contain macroscopically distinct divarications for a moving particle.

\subsection{Conclusion}

So, by our approach, quantum nonlocality of entangled states is an artifact of the
existing quantum theory. It appears in quantum mechanics due to the inconsistency of
its superposition principle with the corpuscular properties of a particle. This
principle should be corrected.

Namely, (1) it must distinguish, on the conceptual level, macroscopically distinct
states and their superpositions; (2) it must forbid introducing observables for
entangled states; (3) it must require decomposing entangled states into the set of
unentangled ones, when such a set has not yet been found.


\begin{thebibliography}{861}
\bibitem{Ens}
A. Einstein, B. Podolsky, and N. Rosen, Phys. Rev. 47, 777 (1935).
\bibitem{Bel}
J. S. Bell, Physics 1, 195 (1964).
\bibitem{Asp}
A. Aspect, Nature (London) 398, 189 (1999).
\bibitem{Ber}
R. A. Bertlmann and A. Zeilinger (eds.), Quantum [Un]speakables, from Bell to
Quantum Information, Springer Verlag, Heidelberg, (2002).
\bibitem{No1}
T. Norsen, e-prints quant-ph/0404016v2.
\bibitem{No2}
T. Norsen, e-prints quant-ph/0408105v3.
\bibitem{No3}
T. Norsen, e-prints quant-ph/0408178.
\bibitem{Wis}
H. M. Wiseman, e-prints quant-ph/0509061.
\bibitem{Hei}
Werner Heisenberg, The Physical Principles of the Quantum Theory (Dover
Publications, 1949), p. 39.
\bibitem{Bee}
J. S. Bell, Speakable and Unspeakable in Quantum Mechanics, (Cambridge University
\bibitem{Gis}
Ll. Masanes, A. Acin, and N. Gisin, Phys. Rev. A 73, 012112 (2006).
\bibitem{Shi}
A. Shimony, 'Controllable and uncontrollable non-locality', in S. Kamefuchi et al.
(eds.), Foundations of Quantum Mechanics in the Light of New Technology (Physical
Society of Japan, Tokyo, 1984), pp. 225-230.
\bibitem{Hor}
M. Horodecki, P. Horodecki, R. Horodecki, J. Oppenheim, A. Sen(De), U. Sen, and B.
Synak-Radtke, Phys. Rev. A 71, 062307 (2005). Press, Cambridge, 1987).
\bibitem{Ch1}
N. L. Chuprikov, e-prints quant-ph/0602172.
\bibitem{Smo}
Charles H. Bennett, David P. DiVincenzo, Christopher A. Fuchs, Tal Mor, Eric Rains,
Peter W. Shor, John A. Smolin, William K. Wootters, e-prints quant-ph/9804053v4.
\bibitem{Ha2}
E.H. Hauge and J.A. St\o vneng, Rev. Mod. Phys. {\bf 61}, 917 (1989).
\bibitem{La1}
R. Landauer and Th. Martin, Rev. Mod. Phys. {\bf 66}, 217 (1994).
\bibitem{Olk}
V.S. Olkhovsky and E. Recami, Phys. Repts. {\bf 214}, 339 (1992).
\bibitem{Ste}
A.M. Steinberg, Phys. Rev. Let. {\bf 74}, 2405 (1995).
\bibitem{Mu0}
J.G. Muga, C.R. Leavens, Phys. Repts. {\bf 338}, 353 (2000).
\bibitem{Nu0}
C.A.A. Carvalho, H.M. Nussenzveig, Phys. Repts. {\bf 364}, 83 (2002).
\bibitem{Ol3}
V. S. Olkhovsky, E. Recami, J. Jakiel, Phys. Repts. {\bf 398}, 133 (2004).

\bibitem{Ha1}
E.H. Hauge, J.P. Falck and T.A. Fjeldly, Phys. Rev. B {\bf 36}, 4203 (1987).
\bibitem{Ja1}
W. Jaworski and D.M. Wardlaw, Phys. Rev. A {\bf 37}, 2843 (1988).
\bibitem{Le1}
C.R. Leavens and G.C. Aers, Phys. Rev. B {\bf 39}, 1202 (1989).
\bibitem{But}
M. Buttiker, Phys. Rev. B {\bf 27}, 6178 (1983).
\bibitem{Le2}
W.R. McKinnon and C.R. Leavens, Phys. Rev. A {\bf 51}, 2748 (1995).
\end{thebibliography}
\end{document}